\documentclass[11pt,dvips]{article}

\usepackage{amssymb,amsfonts}     
 \usepackage{graphicx}
\usepackage{epsfig}
\usepackage{amsmath}
\usepackage{amsbsy}

\usepackage{mathrsfs}

\begin{document}

\title{
{\Huge{  
{\textbf{Vacuum Structure and Potential}} 
}
}}
\author{J X  Zheng-Johansson
\\
  Institute of Fundamental Physics Research,  611 93 Nyk\"oping, Sweden. 
\\
 March, 2007
} 
\date{}
\maketitle

\def\elsub{\mbox{\small${\sf e}$}}
\def\minus{\mbox{{\rm -}}}
\def\v{{\rm v}}
\def\ev{\epsilonp}
\def\epsilonp{\mbox{{\scriptsize$\in$}}}
\def\Vv{-\hspace{-0.3cm}{V}_{\rm v}}  %
\def\Vo{V_o}
\def\rv{{\mit a}_\v}
\def\F{F}
\def\sang{\sigma}
\def\lsub{{_{\mbox{\scriptsize $l$}}}}
\def\Jc{Z_\Xssub}
\def\Xcal{\chi}
\def\Xc{\Xcal}
\def\vsub{{}_{{\rm v}} }
\def\Xsub{{\mbox{\scriptsize${X}$}}}
\def\Xssub{{\mbox{\tiny${X}$}}}

\begin{abstract}
Based on overall experimental observations, especially the pair processes, I developed a model structure of the vacuum along with a basic-particle formation scheme begun in 2000 (with collaborator P-I Johansson). The model consists in that the vacuum is, briefly, filled of neutral but polarizable vacuuons, consisting each of a p-vaculeon and n- vaculeon of charges $+e$ and $-e$ of zero rest masses but with spin motions, assumed interacting each other with a Coulomb force. The model has been introduced in full in a book (Nova Sci, 2005) and referred to in a number of journal/E-print papers. I outline in this easier accessible paper the detailed derivation of the model and a corresponding  quantitative determination of the vacuuon size. 

\end{abstract}
\def\Unifcite{4}
\def\minus{\mbox{-}}
\def\mpm{\mbox{$+$\hspace{-0.1cm},-}}
\def\vel{v}
\def\phiv{\varphi}
\def\aph{\alpha}
\def\Lamd{{\mit \Lambda}_d{}}
\def\Pfp{\Pm_\vel}
\def\Efp{\Eng_\vel}
\def\imc{\mbox{\scriptsize{vir}}}
\def\Pm{P}
\def\Dc{a_{\Cssub}}
\def\Cssub{{\mbox{\tiny${C}$}}}

\def\Mch{\mathfrak{M}}
\def\pac{\mathscr{Y}}
\def\R{\mathfrak{R}}
\def\beat{{\rm b}}
\def\lb{{\bf l}}
\def\vb{{\bf v}}

\def\Rb{{\bf R}}
\def\pd{\partial}
\def\vphi{\varphi}
\def\psiR{\widetilde{\psi}}
\def\psiL{\widetilde{\psi}^{{\rm vir}}}
\def\PhimR{\widetilde{ {\mit \Phi}}}
\def\PsimR{\widetilde{ {\mit \Psi}}}
\def\PsimL{{\widetilde{ {\mit \Psi}}}^{{\rm vir}}}
\def\a{\alpha}
\def\uav{\bar{u}}
\def\D{\Delta}
\def\th{\theta}
\def\r{{\mbox{\tiny${R}$}}}
\def\re{{\mbox{\tiny${R}$}}}
\def\Fmed{F_{{\rm a.med}}}
\def\med{{\rm med}}
\def\Lw{L_{\varphi}}

\def\Efb{{\bf E}}
\def\Bfb{{\bf B}}
\def\Ac{ \varphi}
\def\Xsub{{\mbox{\tiny${X}$}}}
\def\Xssub{{\mbox{\tiny${X}$}}}

\def\Ksub{{\mbox{\tiny${K}$}}}
\def\W{{{\mit \Omega}}}
\def\Wd{\W_d{}}
\def\Nu{{\cal V}}
\def\Nud{\Nu_d{}}
\def\Eng{{\cal E}}
\def\eng{{\varepsilon}}
\def\Acuni{\Ac_{{\Ksub}^\dagsup}^{\dagsup}}
\def\unduni{\Ac_{{\Ksub}^\dagger}^{\dagsup}}
\def\Acauni{\Ac_{{\Ksub}^\ddagsup}^{\ddagsup}}
\def\Acunim{{\Ac_{{\Ksub}^\dagsup}^{\dagsup *}}}
\def\undunim{{\Ac_{{\Ksub}^\dagsup}^{\dagsup}}^*}
\def\Acaunim{{\Ac_{{\Ksub}^\ddagsup}^{\ddagsup *}}}
\def\pd{\partial}
\def\Ad{ {\mit \psi}}
\def\psim{ {\mit \psi}}
\def\Kd{K_d{}}
\def\Lam{{\mit \Lambda}}
\def\lam{\lambda}
\def\dagsup{{\mbox{\tiny${\dagger}$}}}
\def\ddagsup{{\mbox{\tiny${\ddagger}$}}}
\def\psimKdK{\psim_{\Ksub,\Kdsub}}
\def\w{\omega{}}
\def\wdlow{\omega_d }
\def\g{\gamma{}} 
\def\Phim{{\mit \Phi}}
\def\Psim{{\mit \Psi}}
\def\arm{{\rm a}}
\def\brm{{\rm b}}
\def\crm{{\rm c}}
\def\drm{{\rm d}}
\def\erm{{\rm e}}
\def\frm{{\rm f}}
\def\grm{{\rm g}}
\def\hrm{{\rm h}}
\def\lf{\left}
\def\rt{\right}
\def\Kdsub{{\mbox{\tiny${K_d}$}}}
\def\psimkd{\psim_{\kdsub}}
\def\psimKd{\psim_{\Kdsub}}
\def\hquad{ \ \ } 
\def\Taum{{\mit \Gamma}}
\def\Tssub{{\mbox{\tiny${T}$}}}
\def\Kssub{{\mbox{\tiny${K}$}}}
\def\Xssub{{\mbox{\tiny${X}$}}}
\def\Wssub{{\mbox{\tiny${W}$}}}

\def\el{\mbox{${\sf e}$}}

\section{Introduction}

{\it Vacuum} is the continuum in the absence of all material particles  like the molecules and atoms, and the matters or substances made up of these.  
That this vacuum continuum is itself substantial is uniformly pointed to by a range of  phenomena, especially the  pair processes which take place at the interface between (ordinary) matter and vacuum\cite{Anderson1933,Blackett1933,Dirac1930}. 
I shall here mainly  discuss the  indication by the pair processes,
illustrated for the electron-positron annihilation, and derive based on this a model vacuum structure. This will point to a  vacuuonic vacuum structure whose inner property has not been appreciated prior to the author's recent work (with P-I J). The model structure of the vacuum, and a particle formation scheme along with it, 
 have, in terms of first-principles classical-mechanics solutions, facilitated predictions [\Unifcite a-k] of  a range of  properties of basic particles and the vacuum  that are directly comparable with observations. 

\section{The observational substantial vacuum}\label{sec-2}
When an electron $\el^{\minus}$ meets a positron $\el^+$, 
pair annihilation can occur,  
with a highest probability if both being at rest, in a reaction process: 
$$
\displaylines{\refstepcounter{equation} \label{eq-par} 
\hfill 
\el^{\minus} +\el^+ \rightarrow \gamma+ \gamma  \hfill (\ref{eq-par})
}$$
The observed final products are the two gamma rays, $\gamma$'s on the right-hand side, emitted in opposite directions. 
 These  carry the energies converted from and {\it only} from  the {\it masses} of the electron and positron,  $M_{\elsub^{\minus}} +M_{\elsub^+} \ge 2\times 0.511$ MeV. 
The energy equation for this is: 
\begin{eqnarray}\label{eq-pair2}
  M_{\elsub^{\minus}}c^2+ M_{\elsub^+}c^2=h \Nu_\gamma + h \Nu_\gamma 
\end{eqnarray}
with $c$ the velocity of light, $h$ Planck constant, and $\Nu_\gamma$ the frequency of the emitted gamma rays. 
(\ref{eq-pair2}) is today (incompletely) regarded as a total energy equation for the pair process.

In addition to the masses,  the  electron $\el^{\minus}$ and  positron $\el^+$  on the left-hand side of (\ref{eq-par})
we stress  are described by their another elementary   properties,  the charges $-e$ and $+e$, that  maintain an interaction energy between the two  according to  the Coulomb's law.

To explicitly reflect both the two elementary and independent parameters, the mass and the  charge  of each particle that are each associated with an  energy of a specified form and amount,  we put these as  independent variables, for the entities  
$\el^{\minus}(M_{\elsub^{\minus}}, -e) $, $\el^+(M_{\elsub^+}, +e) $,  appearing on the left-hand side  of  (\ref{eq-par}).
         (Their spins are by observation unchanged 
in the reaction and are thus not explicated here.) 
On an equal footing,  
in order to conserve the energy 
we need to include on the right-hand side of (\ref{eq-par})  two new 
entities, $\v_p $ and $\v_n$, that are requisite as the  carriers of the Coulomb interaction energies between  $e $ and $ -e$, so that  
  (\ref{eq-par}) modifies to the complete form 
\begin{eqnarray}\label{eq-pair2}
\el^{\minus}(M_{\elsub^{\minus}}, -e) +\el^+(M_{\elsub^+}, e)  \rightarrow \gamma  (h \Nu_\g)+ \gamma(h \Nu_\g)  + \v_n (-e)+\v_p(e)
 \end{eqnarray}

We below prove that  the $\v_n (-e)$ and $ \v_p(e)$  must exist on the right-hand side of (\ref{eq-pair2}) in order to conserve energy.  Suppose  an electron and positron annihilate at a
separation distance $a_\v$ in the vacuum and, for simplicity when at rest; assume at the scale of $a_\v$ their charges continue to interact according to Coulomb's law.   (Any specific form of interaction will affect its quantitative scale only but not the general energy conservation equations below.) 
Just before the annihilation,  the two particles 
carry therefore a zero total kinetic energy  
$\Eng_\vel (a_\v)=0$, and a Coulomb interaction potential energy  
$$\displaylines{\refstepcounter{equation} \label{eq-a}
\hfill 
V(a_\v)=-\frac{e^2}{4\pi \ev_0 a_\v}              \hfill (\ref{eq-a})
\cr
\mbox{Or, they together  carry a total mechanical energy   
}\hfill
\cr\refstepcounter{equation} \label{eq-b}
\hfill 
\Eng_{\mbox{$+$\hspace{-0.1cm},-}}
 (a_\v) =V(a_\v) + \Eng_\vel (r_0) =  -\frac{e^2}{4\pi \ev_0 a_\v}+0.       \hfill (\ref{eq-b})
}$$
After the annihlation, $\Eng_{\mpm} (a_\v)$ which cannot be destroyed, must therefore necessarily present in a certain (new or the same) form 
$\Eng'_{\mpm} (a_\v)$ which is assumed to be carried by certain entities  denoted earlier by $\v_p$ and $\v_n$.  
 Suppose  $\v_p$ and $\v_n$, are so chosen that they carry all of the mechanical energy and are subject to no external force (the latter will plausibly  hold if the products are as a whole electrically neutral  and situated in unperturbed vacuum). Then, it follows from the requirement of energy conservation that 
the total energy after annihilation must be   
$$\refstepcounter{equation} \label{eq-c}
\Eng'_{\mpm} (a_\v) \equiv \Eng_{\mpm} (a_\v), 
\eqno (\ref{eq-c})
$$  
with $\Eng_{\mpm} (a_\v) $ given by  (\ref{eq-b}) and being finite, and in fact being enormously large for $a_\v$ extraordinarily small. 
Hence, that the annihilation products  $\v_p $ and $ \v_n$ carry an  energy equal to the mechanical energy  $\Eng_{\mpm} (a_\v) $ of the electron-positron, is proven.

As to the form of $\Eng'_{\mpm} (a_\v) $; formally $\Eng'_{\mpm} =V'+\Eng'_\vel$.
Since no observational indications suggest otherwise,  it would be natural   to assume in the first place that after the annilation,
the two charges $-e $ and $ +e$ are unchanged in charge quantities but only transformed to  new dynamical states in the forms of the new entities $\v_n$ and  $\v_p $, 
that are separated  $a_\v$ apart,  
and that they continue to  interact by $V'(a_\v)$ according to  Coulomb's law. 
Then, 
$V'(a_\v)=V(a_\v)=  -\frac{e^2}{4\pi \ev_0 a_\v}$.    
Since $a_\v$ is after annihilation unchanged here, 
 the (center-of-mass) kinetic energy must be   
$\Eng'_\vel(a_\v)=\Eng_\vel(a_\v)=0$. 
($\v_n$ and  $\v_p $ ought to have rotations that preserve the  spins kinetic energies  of the electron and position; see \cite{Unif1} for an explicit treatment.)

\section{The vacuum structure: Proposition}

Two such entities as $\v_n$ and $\v_p$ as probed with today's instruments  are together merely an electrically neutral point in the vacuum. 
That is, such a point is no different from any other points and may apparently be assigned to everywhere in the vacuum. 
Such a point ought therefore necessarily to be assigned to everywhere in a vacuum which gives off pair productions on absorbing a high energy light qunta  and which propagates light wave everywhere the same way. 
We therefore induce that, the vacuum must be filled of such entities, 
$\v_n$ and $\v_p$ paired together, by Coulomb energy $\Vv \equiv V(a_\v)$  at a distance $a_\v$, 
so as to be externally electrically neutral. 
We call each such entity a {\bf vacuuon}.

An electrically neutral vacuuon may be formed of two concentric opposite charged spheres, or alternatively a jelly-like mixture of two intermingling opposite charged fluids. 
The latter, the jelly-like structure, must however be rejected, 
in order to account for the  positron experiments.
These experiments inform  that single positrons do not exist as free particles \cite{Anderson1933,Blackett1933,Dirac1930},
they once created will appear to instantly vanish into the vacuum.  In other terms, the vacuum 
seems to be a negatively charged sea, as originally proposed by Paul Dirac in 1930\cite{Dirac1930}, which traps 
any single positive charge in it, unless the charge has a high enough (oscillatory) kinetic energy (amounting to a proton mass) to escape.

\begin{figure}[htbp]
\vspace{0.2cm}
\centering
\includegraphics[width=0.85\textwidth]{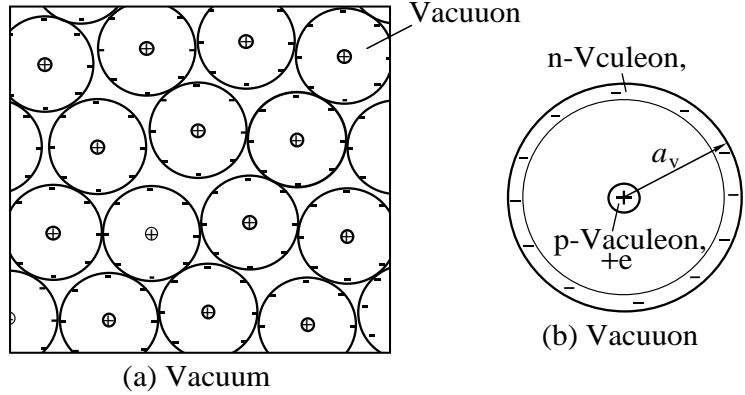}
\caption{
Schematic diagrams of the structures of 
(a)  the  vacuum, being  uniformly filled of electrically neutral vacuuons, and  (b) a vacuuon, being  composed of a p-vaculeon at the center (being point-like), of a charge +e, and an n-vaculeon of a charge -e on a concentric spherical shell  of a radius $\rv$ as envelope. The finite thickness of the n-vaculeon envelope  and the finite size of the p-vaculeon as drawn are only for illustrating the charges locations; their actual sizes are non observables for the presently available information.}
\label{fig-Vacuon}
\end{figure}

Based on the above considerations taking the simplest structure for two concentric spheres, I propose a model for the vacuum structure to be: 

\begin{center}
     \begin{minipage}[t]{12.5cm}
 1. The vacuum is uniformly filled of building entities termed  {\bf vacuuon}s that are electrically neutral and essentially completely at rest in the absence of external disturbances, and  have each a zero rest mass 
(Figure \ref{fig-Vacuon}a).
     \end{minipage}
\end{center}
\vspace{-0.cm}

\begin{center}
     \begin{minipage}[t]{12.5cm}
 \indent
2. A vacuuon is composed of a  {\bf p-vaculeon} having an electric charge $+e$
and 
an {\bf n-vaculeon} having an electric charge $-e$, where $e$ $= 1.602  \cdot 10^{-19}$ C being the elementary charge.  
 The p- and n-vaculeons have each a zero rest mass. 
\indent 
 3. In the vacuuon in its ground-state, the n-vaculeon forms an envelope, a spherical shell of radius $a_\v$, 
about the point-like p-vaculeon at the center (Figure \ref{fig-Vacuon}b). 
\\
\indent 
 4.  The n- and p- vaculeons have each a rotational motion, or spin (a more suitable term in their freed states),  about their own coincidental axes, and hence have a angular momentum, $L_s$. \  $L_s$ equals $\mp \frac{1}{2}\hbar$ as for the  electron and proton spins,
$2 \pi \hbar $ being Planck constant.
\\  
\indent 
 5. The vacuuon, and the n- and p- vaculeons are assumed to obey the basic laws of classical mechanics  under the equivalent conditions (in particular in a vacuum background) under which the basic laws are established.
     \end{minipage}
        \end{center}

The nature of the vacuum as induced  above is similarly, though indirectly, pointed to by the antiproton production (O. Chamberlain, et. al., 1955, 1957; J. Eades, 1999; E. Segre, 1958), and antineutron production (B. Cork, et al., 1960).
The neutral but polarizable vacuuon construction is also indicated by the observed Lamb shift, and the capability of the vacuum to propagate electromagnetic waves like a sound wave in a stretched string, among others. 
The vacuuonic vacuum structure also enables one to comprehend among other the following three prominent experimental phenomena:
(i). The fact that the annihilation of an "existing" electron with a (yet "non-existing") positron takes place most probably within a material substance.  
We now see that, since the positron is trapped in a negative potential well as mentioned above, to lift it to the (vacuum) level at which annihilation occurs, there requires firstly the presence of some external energy supply; this supply may be provided by the  potential field of a proton or of a matrix of protons as  naturally provided by a material substance.   (ii). The fact that there exist  in nature two and only two species long-lived free basic material particles, the proton and the electron. The fact that the proton is much heavier  
by a specified amount than the electron. 
(iii).
From the two only stable simple (termed also basic) material particles, the proton and the electron with a positive and an negative charge respectively, an atom is formed always with the former at the core and the latter at the outskirt.

 The vacuum structure proposed here resembles  characteristically Dirac's vacuum, with the sea of negatively charged n-vaculeon envelopes loosely corresponding to Dirac's sea of negative charges.

\section{Vacuum potential}

Applying Coulomb's law to the charge $+e$ on the point-like p-vaculeon and the charge element $d q$ on the enveloping n-vaculeon,  integrating over the entire surface on the n-vaculeon spherical shell, we obtain the total force the p- vaculeon acting on the n- vaculeon: 
$\F_{pn}=\int_0^{q }  \frac{e d q }{  4 \pi\ev_0 \rv^2 } 
=  \frac{e  }{  4 \pi\ev_0 \rv^2 } 
(\frac{ -e}{4 \pi } )\int_0^{4 \pi}  d \sang$. Or,   
\begin{eqnarray} \label{eq-Fa1} 
\F_{pn} = - \frac{e^2  }{  4 \pi\ev_0 \rv^2 } 
\end{eqnarray} 

\begin{figure}[htbp]
\vspace{0.2cm}
\centering
\includegraphics[width=0.65\textwidth]{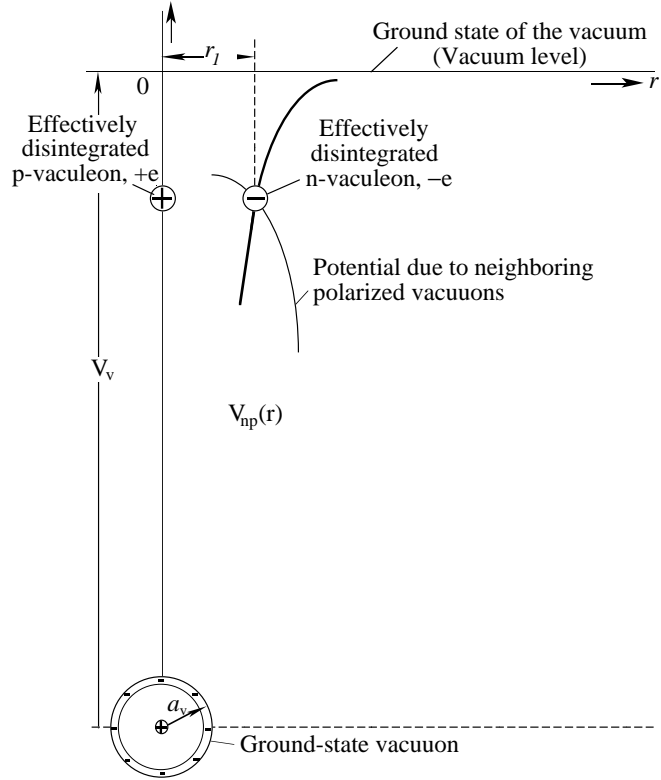}
\caption{
Electrostatic potential $V_{np}(r)$, thick solid curve, between a bound p-vaculeon and n-vaculeon as a function of separation distance $r$, effectively
within $r_1$;  at the equilibrium separation distance $a_\v$ these make up a ground-state vacuuon, 
with $V_{np}(a_\v)=\Vv$. The infinite separation distance corresponds to the vacuum level (vacuum ground state), $V_{np}(\infty)=0$.  As ordinarily situated among other vacuuons close-packed (these making up the vacuum), the two bound vaculeons in the reference vacuuon will be  effectively disintegrated already at a finite distance $<r_1$. Beyond $r_1$ the vaculeon are each primarily subject to the potential field from the surrounding vacuuons as polarized by the vaculeons' own charges.}
\label{fig4.1-pot-vac}\label{figI.1-pot-aether}
\end{figure}

At a variant separation distance $r$, (\ref{eq-Fa1}) writes as $\F_{pn}(r) = - \frac{e^2  }{  4 \pi\ev_0 r^2 }  $.
The interaction potential between the p- and n- vaculeons, or the intervaculeon potential, follows to be
                            \index{Vacuum level, definition of}
$$
V_{np} (r) = - \int_{r}^\infty \F_{pn}d r  =  \int_{\infty}^{r}{ 
  \frac{e^2  }{  4 \pi\ev_0 r^2 } }  d  r = - \frac{e^2  }{  4 \pi\ev_0 r } \eqno({\rm a})
$$
The function $V_{np} (r)$ here corresponds to the $V'(r)$ in Sec. \ref{sec-2}.
Figure \ref{fig4.1-pot-vac} shows a plot of $V_{np} (r)$. 
At the equilibrium separation distance $r=a_\v$, we have a 
{\bf ground-state vacuuon}, and the corresponding potential: 
\begin{eqnarray} \label{eq-Ua} 
\Vv \equiv V_{np}(a_\v)
= - \frac{e^2  }{  4 \pi\ev_0 \rv }
\end{eqnarray}

\vspace{-0.cm}
 The reference level of $V_{np} (r)$,  $V_{np}(r)=0$, is in the above taken at $r\rightarrow \infty$.
The zero intervaculeon potential level $V_{np}(\infty)$
may be taken to coincide with the energy level of {\bf ground-state vacuum}---the vacuum in the absence of any external disturbance and being thus composed of unperturbed neutral vacuuons. Unperturbed neutral vacuuons do not interact with each other and thus have a zero interaction potential.
See discussion in [\Unifcite a]
regarding vacuum potential in the presence of an external vaculeon charge.

$a_\v$ and $V_\v$ have been quantitatively determined with reasonable confidence in [\Unifcite a].  This involves some lengthy procedure, I thus give an outline of the solutions only below. 
Based on classical-mechanics solution for the potential energy of an external p-vaculeon in the vacuum, combined with the minimum kinetic energy requirement for a free p-vaculeon in the vacuum, which is equivalent to a proton mass ($M_p$), $a_\v$ is given by the solution:
$$ \refstepcounter{equation} \label{eq-av0}
a_\v= \frac{-M_pc^2 +\sqrt{(M_pc^2)^2 + 4 (2 \rho_{\lsub}  c^2 )\frac{  \Jc{}' e^2   }{ 4 \pi \ev_0 } }}{2(2 \rho_{\lsub}  c^2) } \eqno(\ref{eq-av0})
$$ 
Where $\Jc{}'$ is a parameter related to the explicit vacuuonic configuration and is estimated (for a dense-packed structure) to be $\sim$1.7; 
$\rho_l$ is the linear mass density of the vacuum. $\rho_l$ can be fixed separately from the  solution for gravitational constant ($G$) [\Unifcite e,b],  to be 
$$\refstepcounter{equation} \label{eq-rhol}
\rho_\lsub 
 = \frac{\Xcal_{0^*} e^4 }{ 4 \pi \epsilonp_0^2 \hbar^2 G}
=1.05(36) \times 10^{-17} \hquad {\rm kg/m}
             \eqno(\ref{eq-rhol})
$$
Where $\Xcal_{0^*} $ is the electric susceptibility of the vacuum and has been more recently determined by an exact solution for between an electron and proton to be (internal work)
$$\refstepcounter{equation} \label{eq-Xcal}
\Xcal_{0^*} =\frac{ \epsilonp_0G M_e M_p}{3e^2} =1.16(9) \times 10^{-41}.
         \eqno(\ref{eq-Xcal})
$$
Using the above $\Jc{}'$ and $\rho_l$ values in (\ref{eq-av0}), $a_\v$ is computed to be:  
$$\refstepcounter{equation} \label{eq-av1}
a_\v\approx 2.70(5) \times 10^{-18} 
\hquad {\rm m} \eqno(\ref{eq-av1})
$$ 
With this $a_\v$ value in (\ref{eq-Ua}), the intervaculeon potential energy of a ground-state vacuuon is estimated to be
$$ 
 \Vv \approx -8.52(8) \times  10^{-11} \hquad {\rm J}
 $$
Or, it equals $-0.532 $ GeV.
Using the $a_\v$ value, the 
the maximum frequency of elastic waves (corresponding to the electromagnetic waves) that can be propagated in the vacuum can be readily estimated to be $\sim 8 \cdot 10^{24}$ 1/s. This is close to the experimentally observed  upper limit of electromagnetic wave frequency.

\section*{Acknowledgements}
The author thanks scientist  P-I. Johansson of Uppsala Univ. for his continued support of the research. 
Prof. R. Lundin has given appreciative review of the  unification scheme conceived and developed by the author, and has  reflected this in the Forward to the two books [\Unifcite a-b] by the author and collaborator P-I. Johansson.

\end{document}